\documentclass[iop]{emulateapj}
\shorttitle{Light curves of Pop III supernovae}
\shortauthors{Tolstov et al.}

\usepackage{color}
\usepackage{enumerate}
\usepackage{amsmath}
\usepackage{textcomp}
\usepackage{bm}
\usepackage[T1]{fontenc}

\begin{document}

\title{Multicolor light curves simulations of Population III core-collapse supernovae: \\ from shock breakout to $^{56}$Co decay}

\author{Alexey Tolstov\altaffilmark{1}, Ken'ichi Nomoto\altaffilmark{1,6}, Nozomu Tominaga\altaffilmark{2,1}, Miho Ishigaki\altaffilmark{1}, Sergey Blinnikov\altaffilmark{3,4,1}, Tomoharu Suzuki\altaffilmark{5}}

\affil{\altaffilmark{1} Kavli Institute for the Physics and Mathematics of the Universe (WPI), The
University of Tokyo Institutes for Advanced Study, The University of Tokyo, 5-1-5 Kashiwanoha, Kashiwa, Chiba 277-8583, Japan} 

\affil{\altaffilmark{2} Department of Physics, Faculty of Science and Engineering, Konan University, 8-9-1 Okamoto, Kobe, Hyogo 658-8501, Japan}

\affil{\altaffilmark{3} Institute for Theoretical and Experimental Physics (ITEP), 117218 Moscow, Russia} 

\affil{\altaffilmark{4} All-Russia Research Institute of Automatics (VNIIA), 127055 Moscow, Russia}

\affil{\altaffilmark{5} College of Engineering, Chubu University, 1200 Matsumoto-cho, Kasugai, Aichi 487-8501, Japan}

\email{$^{*}$ E-mail: alexey.tolstov@ipmu.jp}

\submitted{Accepted for publication in Astrophysical Journal, February 26, 2016}

\begin{abstract}
\noindent
The properties of the first generation of stars and their supernova (SN) explosions remains unknown due to the lack of their actual observations. Recently many transient surveys are conducted and the feasibility of the detection of supernovae (SNe) of Pop III stars is growing. 
In this paper we study the multicolor light curves for a number of metal-free core-collapse SN models (25-100 M$_{\odot}$) to provide the indicators for finding and identification of first generation SNe. We use mixing-fallback supernova explosion models which explain the observed abundance patterns of metal poor stars. Numerical calculations of the multicolor light curves are performed using multigroup radiation hydrodynamic code {\sc stella}. The calculated light curves of metal-free SNe are compared with non-zero metallicity models and several observed SNe.
We have found that the shock breakout characteristics, the evolution of the photosphere's velocity, the luminosity, the duration and color evolution of the plateau -- all the SN phases are helpful to estimate the parameters of SN progenitor: the mass, the radius, the explosion energy and the metallicity.
We conclude that the multicolor light curves can be potentially used to identify first generation SNe in the current (Subaru/HSC) and future transient surveys (LSST, JWST). They are also suitable for identification of the low-metallicity SNe in the nearby Universe (PTF, Pan-STARRS, Gaia).
\end{abstract}



\section{INTRODUCTION}
\label{sec:intro}
\noindent

\footnotetext[6]{Hamamatsu Professor.}

After the big bang, small density fluctuations and gravitational
contraction lead to form first stars, called Population III stars (Pop
III stars). The formation initiates the baryonic evolution of the Universe, e.g.,
formation of first galaxies and cosmic reionization. The formation has
been studied by cosmological simulations for long while
\citep[e.g.][]{BrommYoshida2011}. However, their
nature remains elusive. In particular, the initial mass function of the
first stars, and thus the supernova (SN) explosions of the first stars,
are still exciting issues \citep[e.g.,][]{Hirano2014,Susa2014}.

The nature of the first stars has been mainly studied
with low-mass stars in the Galactic halo. They have a lifetime longer
than the current age of the Universe, and thus preserve chemical abundance at their formation. Such stars are called metal-poor stars. Aiming at the
understanding of the evolution of the early Universe, many surveys of
metal-poor stars have been conducted so far
\citep[e.g.,][]{BeersChristlieb2005} and follow-up high-dispersion
spectroscopic observations have revealed their detailed abundance ratios
\citep[e.g.,][]{Cayrel2004,Yong2013}. The metal-poor stars are
classified with Fe and C abundances, for example, metal-poor
(MP) stars with $\rm{[Fe/H]}<-1$, very metal-poor
(VMP) stars with $\rm{[Fe/H]}<-2$, extremely metal-poor stars (EMP) with
$\rm{[Fe/H]}<-3$, ultra metal-poor stars (UMP) with $\rm{[Fe/H]}<-4$, hyper metal-poor stars (HMP) with $\rm{[Fe/H]}<-5$, and carbon-enhanced metal-poor (CEMP) stars with $\rm{[C/Fe]}>+1$. The observational studies of the metal-poor stars present the larger fraction of CEMP stars at lower [Fe/H]
\citep[e.g.,][]{Hansen2014}. Especially, all of HMP stars show extremely
high C abundance with $\rm{[C/Fe]}>+3$.

\begin{deluxetable*}{lccccccccccc}
\tablecaption{Zero metallicity explosion models\label{modelTable}}
\tablewidth{0pt}
\tablehead{
\colhead{Model} & $Z$ &\colhead{$M$}& $T_{\rm{c}}$ & Luminosity  &\colhead{Radius} & \colhead{$M$(H)} & \colhead{Energy} & \colhead{$M_{\rm{cut}}$(ini)} & \colhead{$M_{\rm{mix}}$(out)} 
& \colhead{$M$($^{56}$Ni)} & \colhead{[C/Fe]$^*$} \\
\colhead{} & &  \colhead{[M$_\odot$]} & \colhead{[$10^3$K]} & \colhead{[$10^6$L$_\odot$]} & \colhead{[R$_\odot$]} & \colhead{[M$_\odot$]} & \colhead{[$E_{\rm{51}}$]} & \colhead{[M$_\odot$]} & \colhead{[M$_\odot$]} &   \colhead{[M$_\odot$]}}   
\startdata
{\sc 25z0E1}    & 0 &  25 & 40  & 0.32 &   30 & 11.1 &  1  & -   &  1.7 & 0.2               & 0.9   \\
{\sc 25z0E1m}   &   &     &     &      &      &      &     & 1.7 &  5.7 & 0 ... $10^{-2}$   &  1.9 \\
{\sc 25z0E10}   &   &     &     &      &      &      & 10  & -   &  1.6 & 0.7   &    0.3        \\
{\sc 25z0E10m}  &   &     &     &      &      &      &     & 1.7 &  6.4 & 0 ... $10^{-1}$  & 0.5 \\
{\sc 40z0E1.3m} & 0 &  40 & 27  & 0.88 &   85 & 15.0 & 1.3 & 2.0 & 12.7 & 0 ... $10^{-1}$  & 0.6 \\
{\sc 40z0E30}   &   &     &     &      &      &      & 30  & -   &  2.0 & 0.8            & 0.4   \\
{\sc 40z0E30m} &   &     &     &      &      &      &     & 2.0 &  5.5 & 0 ... $10^{-1}$ &  1.3 \\
{\sc 40z0E30m2} &   &     &     &      &      &      &     & 2.5 & 14.3 & 0 ... $10^{-1}$ &  0.1 \\                
{\sc 100z0E2m}  & 0 & 100 & 3.5 & 2.2  & 2200 & 27.1 & 2.0 & 2.0 &   40 & 0 ... $10^{0}$  & 0.6 \\                
{\sc 100z0E60}  &   &     &     &      &      &      & 60  & -   &  2.3 & 5.2             & 0.03 \\
{\sc 100z0E60m} &   &     &     &      &      &      &     & 2.3 &   40 & 0 ... $10^{0}$ &  -0.3 \\
\vspace{-0.2cm}
\enddata
\tablecomments{The numbers shown are metallicity, main-sequence mass, color temperature, luminosity, radius,  hydrogen mass, explosion energy, mixing-fallback inner and outer mass, $^{56}$Ni mass, carbon-to-iron ratio \\ $^*$ For mixing-fallback models the value for the models with highest amount of $^{56}$Ni is shown.}
\end{deluxetable*}

\begin{deluxetable*}{lccccccccc}
\tablecaption{Non-zero metallicity explosion models\label{modelTable2}}
\tablewidth{0pt}
\tablehead{
\colhead{Model} & $Z$ &\colhead{$M$}& $T_{\rm{c}}$ & Luminosity  &\colhead{Radius} & \colhead{$M$(H)} & \colhead{Energy} 
& \colhead{$M_{\rm{cut}}$} 
& \colhead{$M$($^{56}$Ni)} \\
\colhead{} & &  \colhead{[M$_\odot$]} & \colhead{[$10^3$K]} & \colhead{[$10^6$L$_\odot$]} & \colhead{[R$_\odot$]} & \colhead{[M$_\odot$]} & \colhead{[$E_{\rm{51}}$]}  
& \colhead{[M$_\odot$]} 
&   \colhead{[M$_\odot$]}}  
\startdata

{\sc 20z-3E1}     & 0.001 &  20 (20)    & 2.6 & 0.12 &  760 & 8.7 &  1   &  1.5 & 0;\,0.4    \\
{\sc 20z002E1}    & 0.02  &  20 (18)   & 2.4 & 0.08 &  800 & 8.2 &  1   &  1.5 & 0;\,0.1    \\
{\sc 25z002E1}    & 0.02  &  25 (22)   & 2.4 & 0.17 & 1200 & 8.7 &  1   &  1.5 & 0;\,0.24   \\
{\sc 25z002E1m}   & 0.02  &   25 (18) &  2.4 & 0.17 & 1200 & 8.7 &  1    &  1.7-5.7$^*$ & 0;\,10$^{-3}$;\,0.1  \\
{\sc 40z002E1}    & 0.02  &  40 (22)   & 2.5 & 0.55 & 1700 & 3.7 &  1   &  10.0 & 0   \\
{\sc 40z002E2}    & 0.02  &  40 (22)   & 2.5 & 0.55 & 1700 & 3.7 &  2   &  1.5 & 0.7   \\
\vspace{-0.2cm}
\enddata
\tablecomments{The numbers shown are metallicity, main-sequence mass (presupernova mass), color temperature, luminosity, radius, hydrogen mass, explosion energy, mass cut, mixing-fallback inner and outer mass, $^{56}$Ni mass. By $^*$ mixing-fallback parameters are indicated: $M_{\rm{cut}}$(ini),  $M_{\rm{mix}}$(out). } 
\end{deluxetable*}

In order to derive properties of first stars from the abundance ratios
of the metal-poor stars, theoretical studies are required \citep[see][for a review]{Nomoto2013}. The
theoretical studies have clarified that the abundance patterns of the
C-normal EMP stars are well reproduced by SN explosions with
main-sequence masses $M_{\rm ms}$ of $<100$ M$_\odot$
\citep{UmedaNomoto2002,Limongi2003,HegerWoosley2010,Tominaga2007a,Tominaga2014}.
It is worth noting that there have been no clear signatures of pair-instability
SNe with $M_{\rm ms}$ of $140-300$ M$_\odot$, although a hint of a star
more massive than 300 M$_\odot$ was found in one metal-poor star with $\rm{[Fe/H]}\sim -2.5$ \citep{Aoki2014}.
For CEMP stars with $s$-process elements and $\rm{[Fe/H]}>-3$, the abundance patterns of most CEMP are explained by the mass transfer from AGB
binary companion \citep[e.g.,][]{Lugaro2012} and binary signatures are
found in their observations \citep{Lucatello2005}. For most of CEMP stars with $\rm{[Fe/H]}<-3$ and HMP stars, the C enhancements require faint SNe which eject such a small mass of $^{56}$Ni as $M$($^{56}{\rm Ni})=10^{-3}-0.01$M$_{\sun}$ for CEMP stars and $M$($^{56}{\rm Ni})<10^{-3}$M$_{\sun}$ for HMP
stars \citep[e.g.,][]{Iwamoto2005,HegerWoosley2010,Tominaga2014,Ishigaki2014}. 
While analogies of faint
SNe for CEMP stars had been detected in the present days
(SN~1997D and SN~1999br, e.g., \citealt{Zampieri2003},  SN~2008ha,
\citealt{Valenti2009}), the ejected $^{56}$Ni masses of the faint SN
models for HMP stars are smaller than those estimated from
light-curve analyses of nearby observed SNe
\citep[e.g.][]{Smartt2009}. This could be due to non-existence of
such a faint SN in the present day or selection effect in the observations of nearby SNe. 

On the other hand, observations of gas clouds by \citet{Fumagalli2011} suggest the presences of metal-free pockets at $z \sim2$, possible Pop III remnants at $z=3.5$ are also observed \citep{Crighton2015}. Metal-free pockets give more possibilities for the identification of SNe of Pop III stars (Pop III SNe). Furthermore, as time-domain astronomy recently attracts attention, many transient surveys are conducted (e.g., CRTS \citealt{Drake2009}, PTF \citealt{Rau2009}, ASSASN
\citealt{Shappee2014}, KISS \citealt{Morokuma2014}) and ongoing/planned ones with 8m-class telescopes, e.g., Subaru/HSC (SHOOT,
\citealt{Tominaga2014ATel}) and LSST (Large Synoptic Survey Telescope). The feasibility of the detection of Pop III SNe is growing. Therefore
in order to distinguish them from SNe of Population I or II
stars, their realistic predictions based on the theoretical
models and observations of the metal-poor stars are required.

In this paper we present the light curves for a number of Pop III
core-collapse SNe. Unlike similar numerical simulations performed
by \citet{Whalen2013, Smidt2014}, we mostly concentrate on realistic SN
models, existence of which is indicated by the observed abundance
patterns of metal-poor stars
\citep{Ishigaki2014}. Our accurate consideration of radiative transfer
allows to calculate detailed shape of multicolor
light curves from the shock breakout epoch through the $^{56}$Co decay. The detailed simulations of the light curves can be crucial for detection and
identification of Pop III SNe and this study provides fine points for
the current and future surveys. 

\begin{figure*}
\centering
\includegraphics[width=150mm]{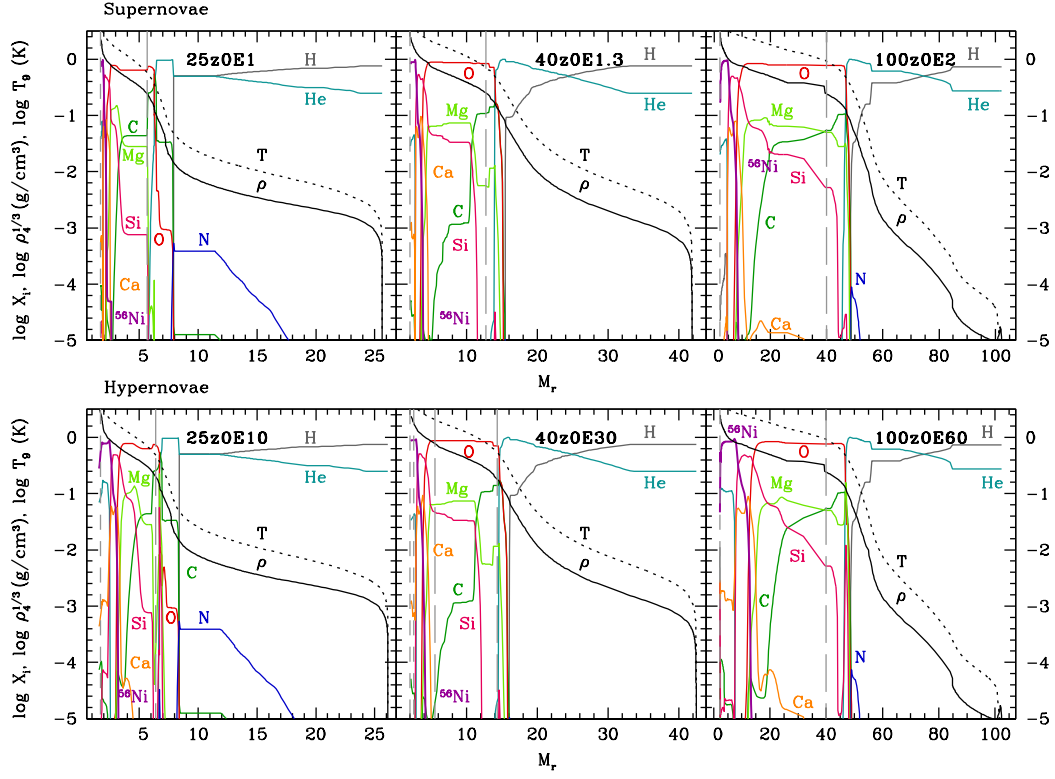}
\caption{The density, temperature of zero metallicity presupernovae and composition after explosive nucleosynthesis. The short dashed and long dashed lines correspond to $M_{\rm{cut}}$(ini) and $M_{\rm{mix}}$(out) in mixing-fallback models.}
\label{chemZ0}
\end{figure*} 
 
\begin{figure*}
\begin{center}
\includegraphics[width=100mm]{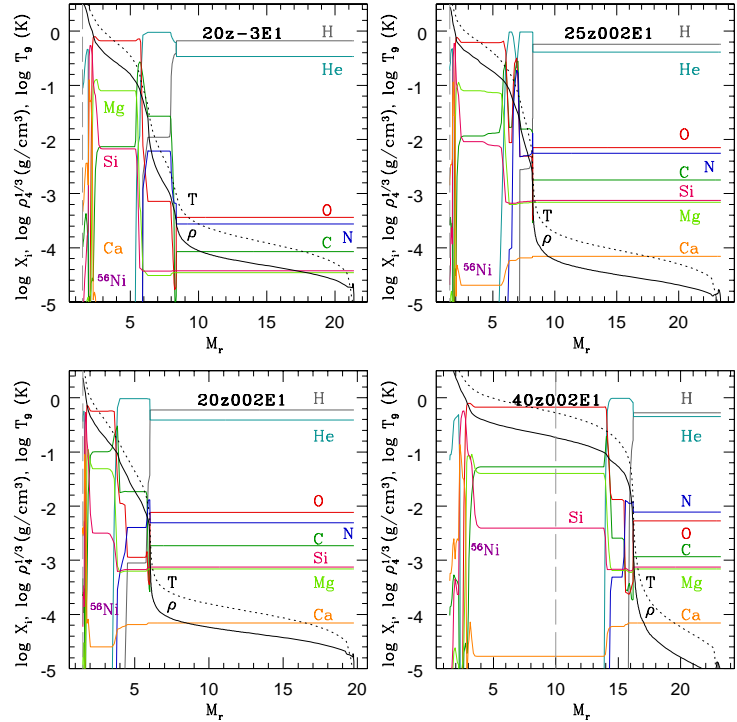}
\caption{The density, temperature of non-zero metallicity presupernovae and composition after explosive nucleosynthesis. Mass cut is shown by long dashed line.} 
\label{chemRSG}
\end{center}
\end{figure*}  

The results of our simulations can also be used for finding and identification of low metallicity SNe.

The paper is organized as follows. In section 2
we describe the mixing-fallback models and numerical methods used in calculations. Section 3 presents the results of the calculations of the light curves from the shock breakout epoch to the $^{56}$Co decay.
In Section 4 we make the comparison of the modeling with observed massive type II SNe. Finally, in Section 5 summary and discussion are given.

\section{MODELS AND METHODS}
\label{sec:models}
\subsection{Models}
\noindent

We calculate light curves for zero metallicity progenitors
with the main-sequence masses $M_{\rm{MS}}=25,40,100\,$M$_{\odot}$ and the explosion energies corresponding to SNe ($E_{\rm{51}}\equiv E/10^{51}$ erg = 1) and hypernovae (HNe)($E_{\rm{51}} \geq$ 10) (see Table \ref{modelTable} for details). The presupernova models include detailed postprocess nucleosynthesis calculations \citep{Tominaga2007b}. 
The mass loss is considered as a function of metallicity and is supposed to be zero for Pop III models.
The composition of the models after explosive nucleosynthesis and density structures of presupernova stars are shown in Figure \ref{chemZ0}. Presupernova stars with $M_{\rm{MS}}=25,40$ M$_\odot$ are blue supergiants (BSGs) and these progenitor models have been used by \citet{Ishigaki2014} to find mixing-fallback best-fit models. The transition between red supergiant (RSG) and BSG zero-metallicity stars is still under investigation and we add a RSG progenitor model with $M_{\rm{MS}}=100$ M$_{\odot}$ model, RSG progenitor, to give a more complete picture. The fits of the initial mass distribution to the ensemble of inferred Population III star gives $87^{+13}_{-33}$ M$_{\odot}$ for the maximum progenitor mass \citep{FraserCasey2015}, close to our choice of 100 M$_{\odot}$ model. For the 100 M$_{\odot}$ model we did not find a good mixing-fallback parameters to fit the observed metal poor stars, nevertheless the best models  have a small mass of $^{56}$Ni: $M$($^{56}$Ni$)\sim10^{-7}$ M$_\odot$.

\begin{deluxetable*}{lccccccc}
\tablecaption{Zero metallicity supernovae properties \label{modelTableOutput}}
\tablewidth{0pt}
\tablehead{
\colhead{Model} & \colhead{$\Delta t$} & \colhead{$L$} & \colhead{$u_{\rm{ph}}$} & \colhead{$L_{\rm{SBO}}$} & \colhead{$T_{\rm{eff,SBO}}$} & \colhead{$u_{\rm{ph,SBO}}$} & \colhead{log($N_{\rm{UV}}$)} \\
\colhead{} & \colhead{[days]} & \colhead{[$10^9$L$_\odot$]} & \colhead{[$10^3$km s$^{-1}$]} & \colhead{[$10^9$L$_\odot$]} & \colhead{[$10^6$K]} & \colhead{[$10^3$km s$^{-1}$]} & $>13.6$eV }  
\startdata

{\sc 25z0E1}    &  50 &  0.06 & 1.5 &  6  & 0.3  & 17 & 57 (57) \\
{\sc 25z0E1m}   &  50 &  0.06 & 1.5 &  8  & 0.3  & 17 & 57 (57) \\
{\sc 25z0E10}   &  40 &  0.3  & 5   & 50  & 0.5  & 43 & 59 (58) \\
{\sc 25z0E10m}  &  35 &  0.4  & 5   & 50  & 0.5  & 43 & 57 (57) \\
{\sc 40z0E1.3m} &  50 &  0.1  & 3   & 20  & 0.2  & 15 & 57 (57) \\
{\sc 40z0E30 }  &  35 &  0.9  & 7   & 300 & 0.4  & 53 & 59 (58) \\
{\sc 40z0E30m} &  30 &  1.0  & 7   & 300 & 0.4  & 53 & 59 (58) \\
{\sc 40z0E30m2} &  30 &  1.2  & 8   & 300 & 0.4  & 53 & 59 (58) \\                
{\sc 100z0E2m}  & 135 &  1.5  & 3   &  30 & 0.05 & 3  & 59 (59) \\                
{\sc 100z0E60}  &  70 & 23.3  & 16  & 300 & 0.1  & 19 & 60 (60) \\
{\sc 100z0E60m} &  70 & 34.8  & 22  & 300 & 0.1  & 19 & 60 (60) \\

\vspace{-0.2cm}

\enddata
\tablecomments{The numbers shown are the duration of light curve plateau, the luminosity at the midpoint of the plateau, the photosphere velocity at the mid-plateau epoch, the peak luminosity at shock breakout epoch, photosphere velocity at shock breakout epoch, the number of UV photons in the models with $^{56}$Ni (in brackets the number of UV photons in the models with $M$($^{56}$Ni)=0). Plateau characterisics ($\Delta t$, $L$, $u_{\rm{ph}}$) are shown for models with $M$($^{56}$Ni)=0.}
\end{deluxetable*}

\begin{deluxetable*}{lccccccc}
\tablecaption{Non zero metallicity supernovae properties \label{modelTableOutput2}}
\tablewidth{0pt}
\tablehead{
\colhead{Model} & \colhead{$\Delta t$} & \colhead{$L$} & \colhead{$u_{\rm{ph}}$} & \colhead{$L_{\rm{SBO}}$} & \colhead{$T_{\rm{eff,SBO}}$} & \colhead{$u_{\rm{ph,SBO}}$} & \colhead{log(${N_{\rm{UV}}}$)} \\
\colhead{} & \colhead{[days]} & \colhead{[$10^9$L$_\odot$]} & \colhead{[$10^3$km s$^{-1}$]} & \colhead{[$10^9$L$_\odot$]} & \colhead{[$10^6$K]} & \colhead{[$10^3$km s$^{-1}$]} & $>13.6$eV }  
\startdata

{\sc 20z-3E1}   & 105 & 0.47 & 4   & 15  & 0.07 & 5 & 58 (58) \\
{\sc 20z002E1}  & 115 & 0.47 & 4   & 17  & 0.07 & 5 & 58 (58) \\
{\sc 25z002E1}  & 125 & 0.62 & 3.5 & 35  & 0.07 & 4 & 59 (59) \\
{\sc 25z002E1m} & 135 & 0.74 & 4   & 40  & 0.07 & 4 & 59 (59) \\
{\sc 40z002E1}  & 80  & 0.8  & 4.5 & 35  & 0.06 & 3 & 59 (59) \\
{\sc 40z002E2}  & 70  & 1.7  & 9   & 100 & 0.08 & 4 & 59 (59) \\
\vspace{-0.2cm}

\enddata
\tablecomments{The numbers shown are the duration of light curve plateau, the luminosity at the midpoint of the plateau, the photosphere velocity at the mid-plateau epoch, the peak luminosity at shock breakout epoch, photosphere velocity at shock breakout epoch, the number of UV photons in the models with $^{56}$Ni (in brackets the number of UV photons in the models with $M$($^{56}$Ni)=0). Plateau characterisics ($\Delta t$, $L$, $u_{\rm{ph}}$) are shown for models with $M$($^{56}$Ni)=0.}
\end{deluxetable*}

The aspherical effects are taken into account with the mixing-fallback model which uses three parameters: the initial mass cut $M_{\rm{cut}}$(ini),
the outer boundary $M_{\rm{mix}}$(out) and the ejection factor $f$ \citep{Tominaga2007a}. The parameters of mixing-fallback model are choosen around the values which provide
best-fit to the observed elemental abundances \citep{Ishigaki2014}.
However, in order to investigate the dependence of light curves on the ejected mass of $^{56}$Ni, we vary the ejection factor $f$ from the ordinary supernova
values $0.07-0.2\, $M$_{\odot}$ down to the extremely metal poor case of $M$($^{56}$Ni$)=10^{-7} \, $M$_{\odot}$.

For the 40 and 100 M$_\odot$ SN models we have to slightly increase the explosion energy from $E_{51}$=1 to 1.3 and 2 correspondingly (see Table 1), in order to
avoid the numerical oscillation of internal zones due to the fallback
in the spherical symmetric approximation. 
Thus, the luminosities of these models are higher than those for the originally supposed explosion energy $E_{51}$=1. Estimations for plateau phase (distinctive flat stretch during the decline) gives the increase in 0.3 mag for the 25 M$_\odot$ model and in 0.7 mag for the 100 M$_\odot$ model. As the nucleosynthetic products are not influenced much by
the small change in the explosion energy, these changes do
not lead to any inconsistency.
This fallback issue should be investigated in the future with the use of multi-dimensional calculations.

In addition to zero metallicity progenitors we adopt progenitors with various metallicities up to solar (see Figure \ref{chemRSG} and Table \ref{modelTable2}). The solar metallicity models are RSGs, so that the density and temperature inside the H-rich envelope are about several orders of magnitude lower than those of zero metallicity models.

The solar metallicity models undergo mass loss (Figure \ref{chemRSG}), which is most significant for the 40 M$_{\odot}$ model. We do not apply the mixing-fallback scenario for these models using them only for qualitative comparison with zero metallicity models. The mass of the mixing zone usually does not produce a significant changes of light curves. However, in order to investigate mixing effects, we apply mixing-fallback model to both the 25 M$_{\odot}$ zero and solar metallicity progenitors.

Similar to zero metallicity models to avoid numerical oscillations we have to slightly change the explosion parameters. We adopt 
the explosion energy larger than $E_{\rm{51}}=$ 1 for the 40 M$_\odot$ solar metallicity supernova model {\sc 40z002E2} with low mass cut.

\subsection{{\sc stella} code}

For calculation of the light curves we use the multigroup radiation hydrodynamics numerical code {\sc stella} \citep{Blinnikov1998,Blinnikov2000,Blinnikov2006} and for hypernova simulations we include special relativistic corrections in the hydro code in the manner of \citet{MisnerSharp1969}. {\sc stella} solves implicitly time-dependent equations for the angular moments of intensity averaged over fixed frequency bands and computes variable Eddington factors that fully take into account scattering and redshifts for each frequency group in each mass zone. Here we set $200$ frequency groups in the range from $10^{-3}$ \AA\, to $5\times10^{4}$ \AA. The explosion is initialized as a thermal bomb just above the mass cut, producing a shock wave that propagates outward. The effect of line opacity is treated as an expansion opacity according to the prescription of \citet{EastmanPinto1993} (see also \citet{Blinnikov1998}). The opacity table includes $1.5 \times 10^5$ spectral lines from \citet{KuruczBell1995} and \citet{Verner1996}.

\section{RESULTS}
\label{sec:results}
\noindent

The results of the light curves calculations for zero metallicity models are summarized in Table \ref{modelTableOutput}, for non-zero metallicity models in Table \ref{modelTableOutput2}. Below we discribe in details the results of calculations for zero metallicity models at every phases: from shock breakout through the $^{56}$Co decay. We also highlight the modes reproducing the C-normal EMP ([C/Fe]$\lesssim+1$), CEMP ([C/Fe]$\gtrsim+1$) and HMP ([C/Fe]$\gtrsim+3$) stars and make a comparison with non-zero metallicity SNe.

\begin{figure*}
\begin{center}
\includegraphics[width=80mm]{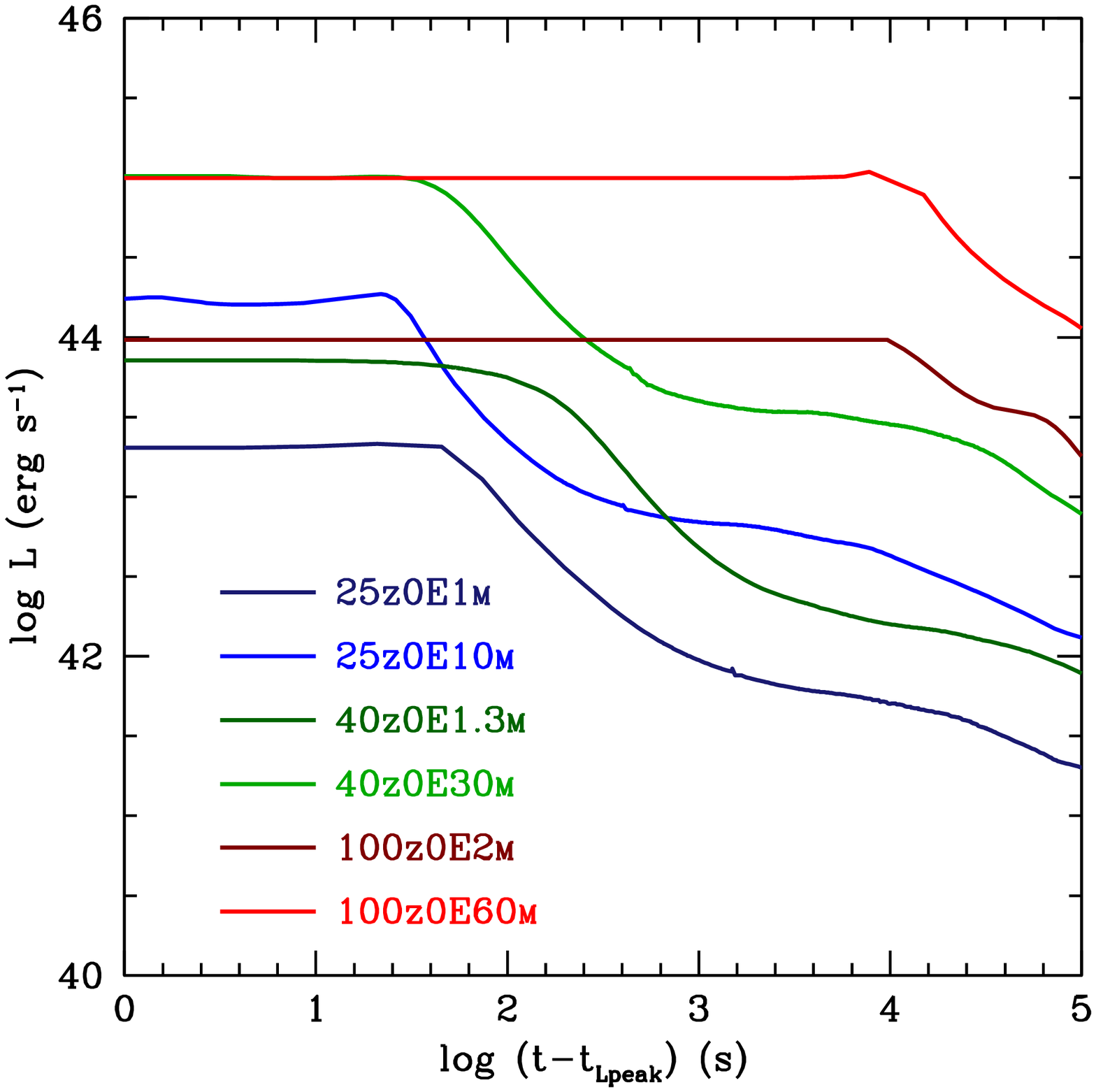}
\includegraphics[width=80mm]{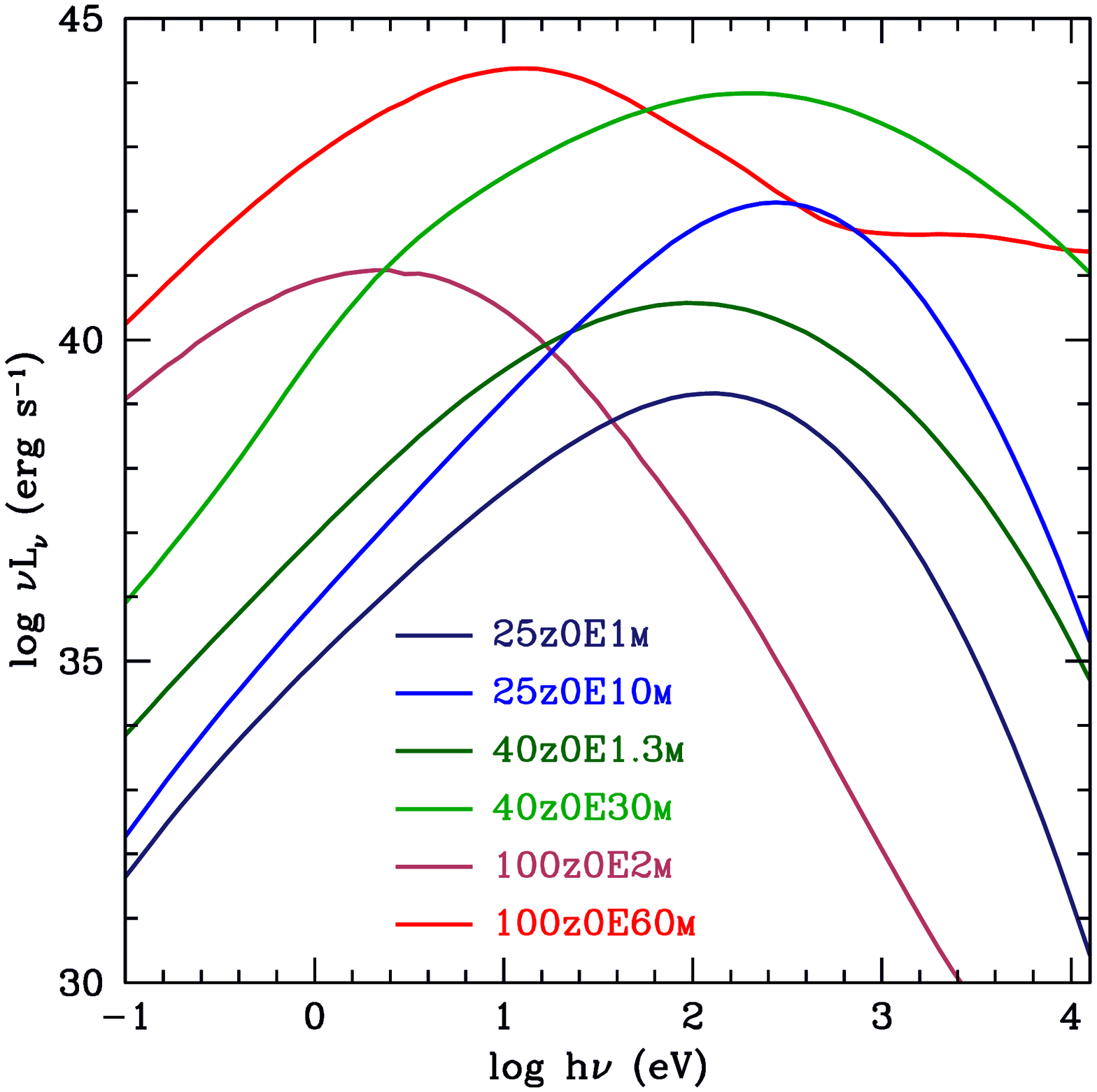}
\caption{Bolometric light curves and spectra at shock breakout epoch for zero metallicity models.} 
\label{sbo}
\end{center}
\end{figure*} 

\begin{figure*}
\centering
  \begin{minipage}[htp]{0.48\textwidth}
\centering
\includegraphics[width=80mm]{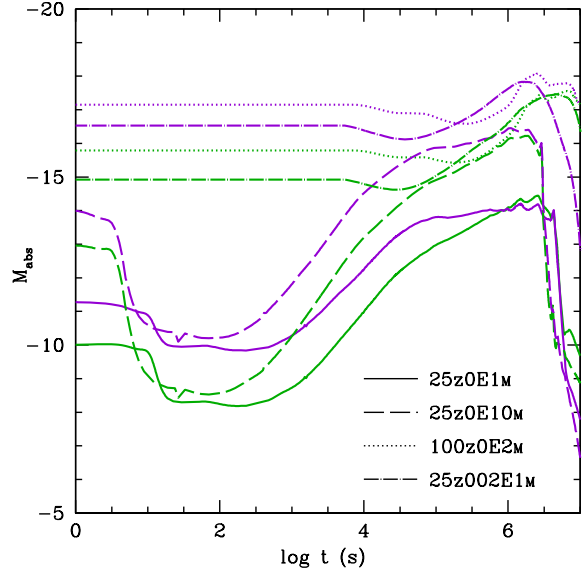}
\caption{U-band (violet color) and V-band (green color) light curves for zero and solar metallicity models from shock breakout to plateau phase.}
\label{faintCRT}
\end{minipage}
\end{figure*}

\begin{figure*}
\includegraphics[width=170mm]{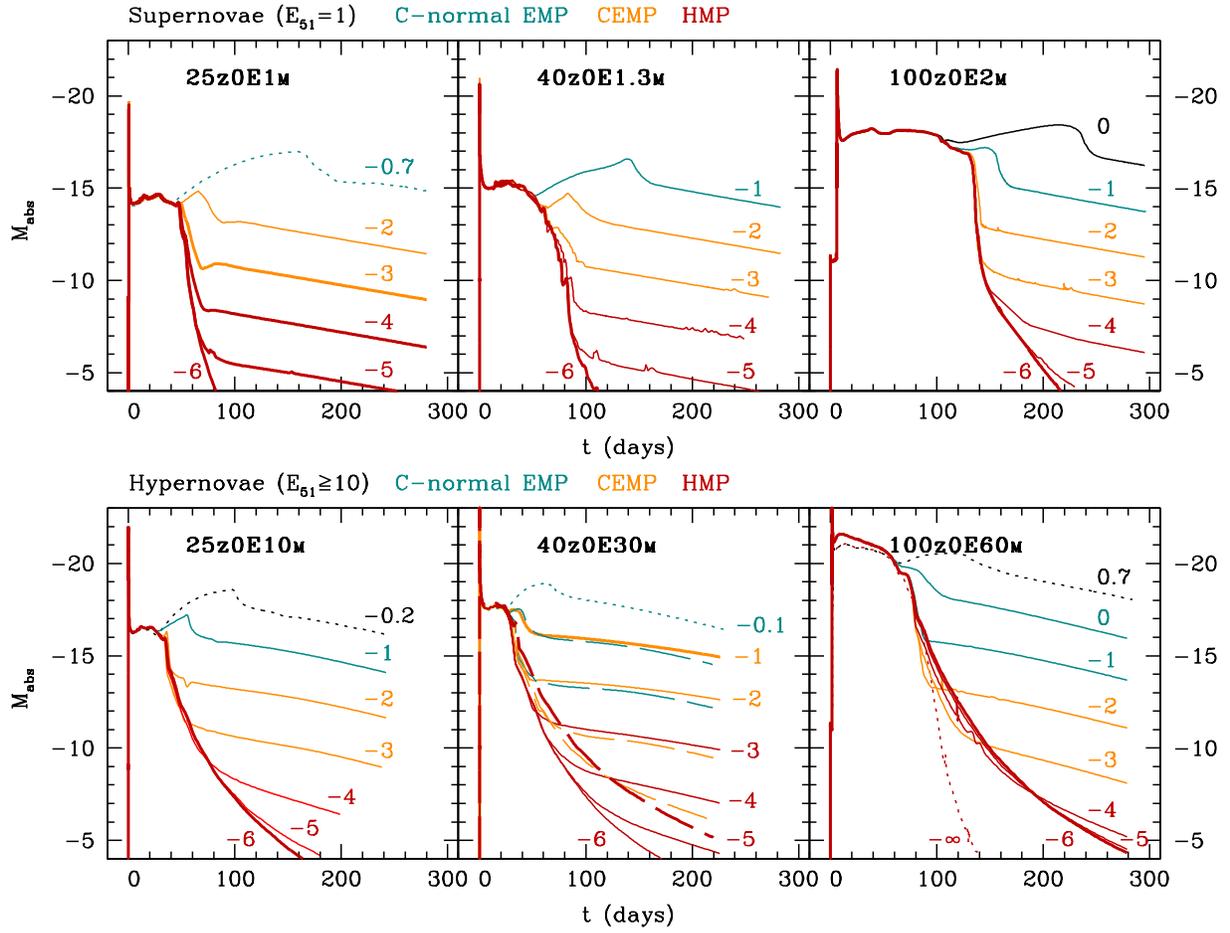}
\caption{Bolometric light curves of zero-metallicity SNe parametrized by log($M$($^{56}$Ni)/M$\odot$) for mixing-fallback models (solid line), original non-mixed models (dotted line) and large mass cut {\sc 40z0E30M2} model (dashed line). Solid lines denote the models with the best fit to metal poor stars. Colors denote the EMP (turquoise color), CEMP (orange color) and HMP models (red color).}
\label{z0bolometric}
\end{figure*}  

\begin{figure}
\centering
  \begin{minipage}[htp]{0.48\textwidth}
\centering
\includegraphics[width=80mm]{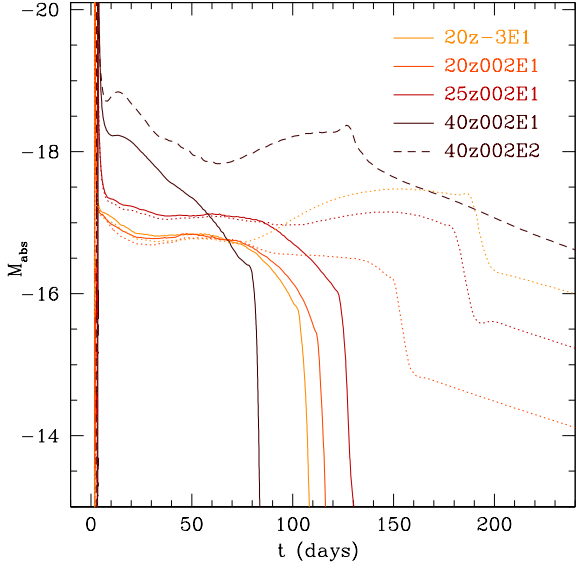}
\caption{Bolometric light curves for non-zero metallicity (close to solar metallicity) progenitor models with zero amount of $^{56}$Ni (solid line) and original non-mixed model (dotted line). Dashed line represents the model {\sc 40z002E2}. In contrast to the model {\sc 40z002E1}, {\sc 40z002E2} has higher explosion energy $\bm{E_{\rm{51}}=}$ 2, lower mass-cut $\bm{M_{cut}}$=1.5M$\bm{_{\odot}}$ and non-zero nickel mass $\bm{M}$($\bm{^{56}}$Ni)=0.7M$\bm{_{\odot}}$. }
\label{rsg}
\end{minipage}
\end{figure}

\begin{figure*}
\includegraphics[width=160mm]{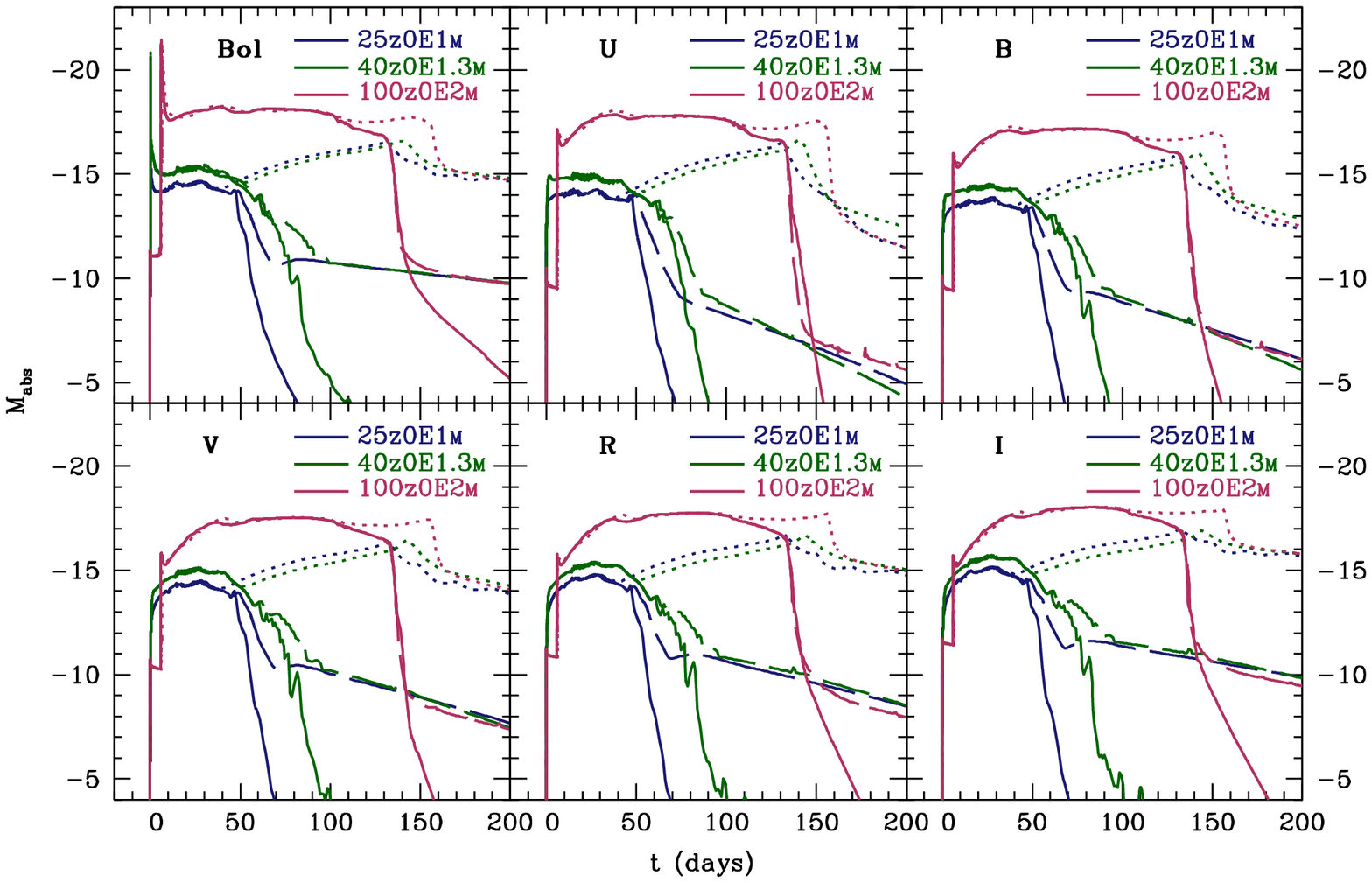}
\caption{Bolometric and UBVRI light curves for zero metallicity mixing-fallback supernova models (the explosion energy $E_{51} \sim$ 1). Solid line -- $M$($^{56}$Ni)=0 (HMP), dashed line -- $M$($^{56}$Ni)=$10^{-3} $M$_{\odot}$ (CEMP), dotted line -- $M$($^{56}$Ni)=$10^{-1} $M$_{\odot}$ (C-normal EMP).}
\label{ubvri}
\end{figure*}   

\begin{figure*}
\includegraphics[width=160mm]{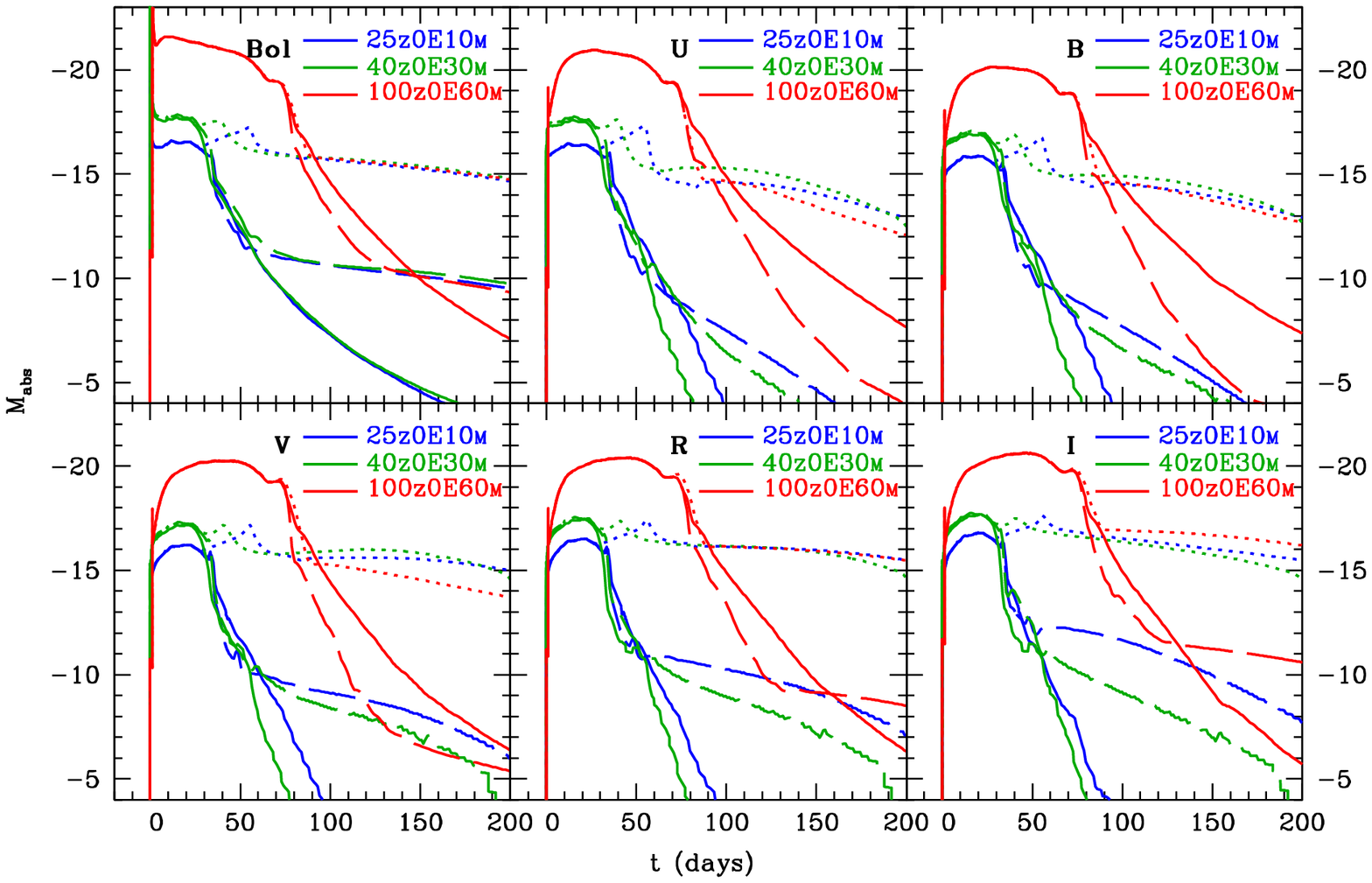}
\caption{Bolometric and UBVRI light curves for zero metallicity mixing-fallback hypernova models (the explosion energy $E_{51} \bm{\geq$} 10). Solid line -- $M$($^{56}$Ni)=0 (HMP), dashed line -- $M$($^{56}$Ni)=$10^{-3} $M$_{\odot}$ (CEMP), dotted line -- $M$($^{56}$Ni)=$10^{-1} $M$_{\odot}$ (C-normal EMP). }
\label{ubvri2}
\end{figure*}

\begin{figure}
\centering
  \begin{minipage}[htp]{0.48\textwidth}
\centering
\includegraphics[width=80mm]{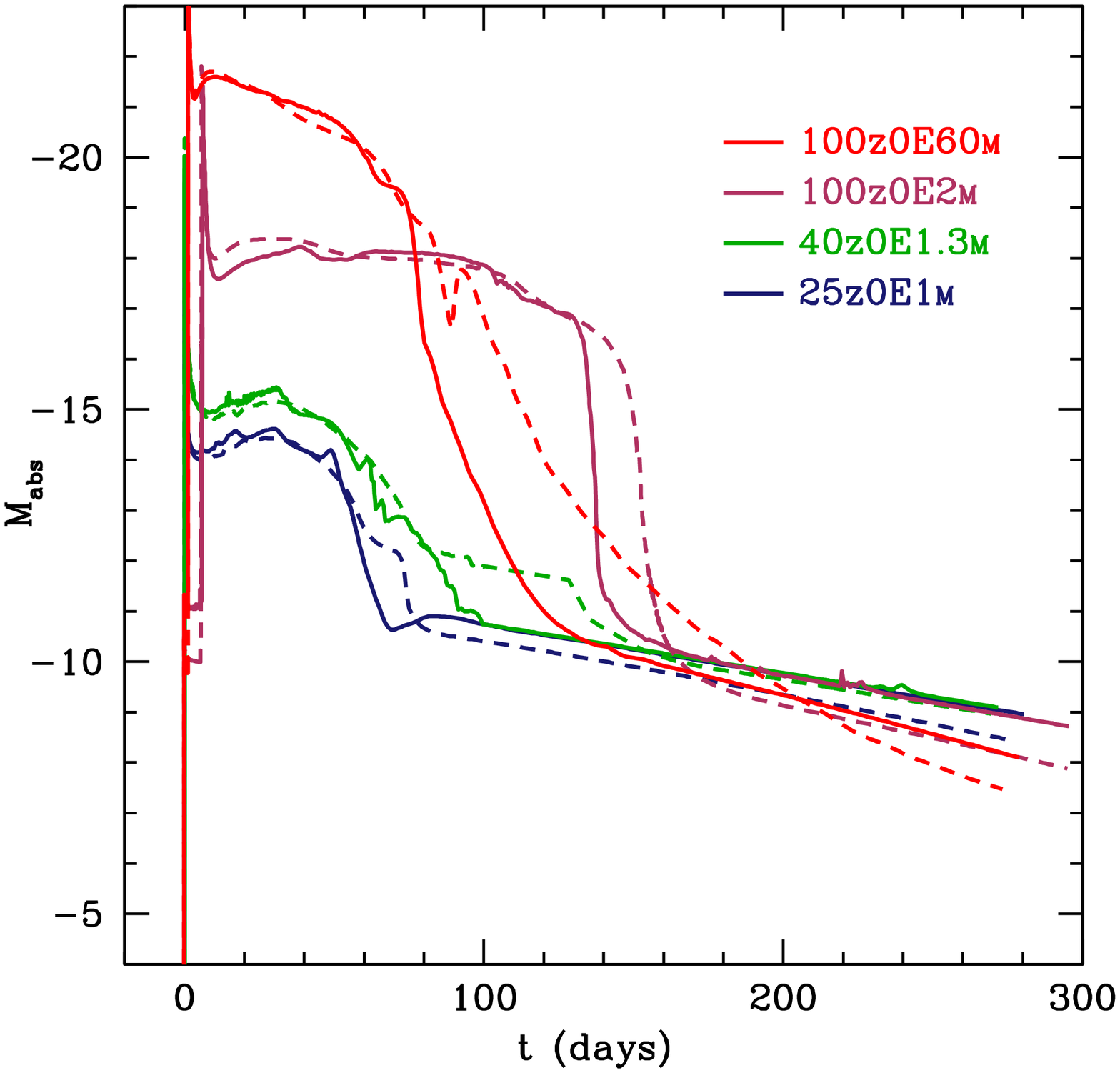}
\caption{Bolometric light curves for zero metallicity progenitor models with $M$($^{56}$Ni)=$10^{-3}$ M$_{\odot}$ (solid line) and mixed model (dashed line).} 
\label{mix}
\end{minipage}
\end{figure} 

\subsection{Shock breakout}

In this subsection we mostly describe the properties of zero-metallicity models at the epoch of shock breakout and the difference between zero and solar metallicity models. More detailed investigation of Type II plateau supernova (SN~II-P) shock breakout for non zero metallicity progenitors has been done by \citet{Tominaga2011}. 

The light curves and spectra at the epoch of shock breakout for zero-metallicity models are presented in Figure \ref{sbo}. The duration of the shock breakout are mostly defined by the radius of the progenitors and varies from $\sim$ 100 s for BSGs up to $\sim$ 1000 s for RSGs. 
The peak frequency of BSG models (25,40 M$_{\odot}$) is in X-ray and their shock breakout can be detected for example by SWIFT/XRT telescope up to $z \sim 0.1$ for SN and $z \sim 1$ for HN models.

The shape of the spectrum for the 100 M$_{\odot}$ hypernova model is different from the others because of several density peaks in the outer layers of the ejecta. These peaks are formed due to inefficient gas acceleration \citep{Tolstov2013} and for such cases more reliable calculations can be done only in multidimensional consideration of  fragmentation in the outer layer.

The shock breakout phase of solar metallicity supernovae has larger luminosity, but the effective temperature is much lower for the same mass in comparison with zero metallicity supernovae (see Table \ref{modelTableOutput2}). The main properties of shock breakouts depend on the radius, the ejecta mass, the explosion energy, the opacity and the density structure of the external layers \citep[see, e.g., analityc estimations of][]{ImshennikNadezhin1988, MatznerMcKee1999}. The dependence of the duration of the shock breakout $\Delta t_{SBO}$ and the peak luminosity $L_{SBO}$ on presupernova parameters can be estimated  as follows \citep{MatznerMcKee1999}:
\begin{align}
 \Delta t_{SBO} =  790 \Big(\frac{\kappa}{0.34\rm{cm}^2\rm{g}^{-1}} \Big)^{-0.58} \Big(\frac{E}{10^{51}\rm{erg}}\Big)^{-0.79} \nonumber \\
\times  \Big(\frac{\rho_1}{\rho_{*}}\Big)^{-0.28} 
\Big(\frac{M}{10\rm{M}_{\odot}}\Big)^{0.21} 
       \Big(\frac{R}{500\rm{R}_{\odot}}\Big)^{-0.79}
        \,\rm{s} \,\, \rm{(RSG)} \,, \nonumber \\ 
 \Delta t_{SBO} =  40 \Big(\frac{\kappa}{0.34\rm{cm}^2\rm{g}^{-1}} \Big)^{-0.45} \Big(\frac{E}{10^{51}\rm{erg}}\Big)^{-0.72} \nonumber \\
\times  \Big(\frac{\rho_1}{\rho_{*}}\Big)^{-0.18} 
       \Big(\frac{M}{10\rm{M}_{\odot}}\Big)^{0.27} 
       \Big(\frac{R}{50\rm{R}_{\odot}}\Big)^{1.90}
        \,\rm{s}\,\, \rm{(BSG)} \,,     
\label{sbo1}
\end{align}
\begin{align}
& L_{SBO} =  2.2\cdot10^{45} \Big(\frac{\kappa}{0.34\rm{cm}^2\rm{g}^{-1}} \Big)^{-0.29} \Big(\frac{E}{10^{51}\rm{erg}}\Big)^{1.35} \nonumber \\
&\times   \Big(\frac{\rho_1}{\rho_{*}}\Big)^{-0.37} 
\Big(\frac{M}{10\rm{M}_{\odot}}\Big)^{-0.65} 
       \Big(\frac{R}{500\rm{R}_{\odot}}\Big)^{-0.42}
        \,\rm{erg\,s^{-1}} \,\, \rm{(RSG)} \,,  \nonumber \\ 
& L_{SBO} =  1.9\cdot10^{45} \Big(\frac{\kappa}{0.34\rm{cm}^2\rm{g}^{-1}} \Big)^{-0.39}  \nonumber \\
& \quad\quad \times \Big(\frac{E}{10^{51}\rm{erg}}\Big)^{1.30} 
  \Big(\frac{\rho_1}{\rho_{*}}\Big)^{-0.23}
 \nonumber \\
& \quad\quad \times       \Big(\frac{M}{10\rm{M}_{\odot}}\Big)^{-0.69}        
       \Big(\frac{R}{50\rm{R}_{\odot}}\Big)^{-0.22}
        \,\rm{erg\,s^{-1}}\,\, \rm{(BSG)} \,,   
\label{sbo2}
\end{align} 
where $\kappa,E,M$ and $R$ are the opacity, the explosion energy, the mass of the ejecta and the radius of the presupernova, respectively. The density factor $\rho_1/\rho_{*}\approx 1$ for RSG models. For BSGs $\rho_1/\rho_{*}$ depends on the density structure, composition and luminosity of the outer layers of the star and varies from $\rho_1/\rho_{*}\approx 50$ for the 25 M$_{\odot}$ model to $\rho_1/\rho_{*}\approx 2$ for the 40 M$_{\odot}$ model.

The difference between RSG and BSG models are clearly seen by the duration of the peak: the larger presupernova radius leads to a longer duration of the shock breakout epoch. In comparison with zero metallicity BSGs, the larger luminosity of solar metallicity SNe is due to lower (by several orders of magnitude) opacity of the external layers of RSG.  The photosphere's temperature of RSG models is only 3,000-4,000 K (in contrast to $\sim$ 20,000 K in BSG models) and the opacity of neutral atoms of hydrogen is dominated by bound-bound and bound-free transitions. In accordance with theoretical estimations this low opacity of exteral layers increases the luminosity of the shock breakout.

\subsection{Cooling phase and rising time}

The shock breakout is followed by "cooling envelope phase" the decline of the luminosity (see Figure \ref{faintCRT}). The duration of this phase strongly depends on the presupernova radius. For compact BSG progenitors the duration is much shorter in optical U and V bands ($\sim$ 100 s) than for RSGs ($\sim$ 1 day). After reaching the luminosity minimum, the light curve starts rising, as the heated expanding stellar envelope diffuses out. For RSG progenitors the rise-time ($\sim$ 10 days) is the order of magnitude longer than for BSG progenitors. This behavior is consistent with the previous investigation of the SN light curves for the early phases \citep[see, e.g.,][]{Gonzalez2015}.

\subsection{Plateau phase}

The presence of the massive hydrogen envelope in zero metalicity presupernova models should produce light curve similar to SNe II-P. The duration of the plateau is determined primarily by the mass of the envelope $M_{\rm{ej}}$ and the other main outburst properties, the explosion energy $E$ and the initial radius $R$ \citep{Litvinova1985,Popov1993}. The SN II-P light curve tails are believed to be powered by the $^{56}$Co decay and the temporal behavior is determined by the ejected mass of $^{56}$Ni \citep[see, e.g.,][]{Nadyozhin1994}. 
The variation of the ejected mass of $^{56}$Ni allows to cover the uncertainty
of matter mixing and fallback in aspherical, jet-like explosion.
We start with $M$($^{56}$Ni) $\sim$  $0.1 $M$_{\odot}$ and logarithmically decrease down to $10^{-6} $M$_{\odot}$ by a factor of $10 $M$_{\odot}$. Lower values of $M$($^{56}$Ni) leads to extremely small changes in the light curve ($M_{\rm{abs}}<-5$) and for practical purposes the light curve with $M$($^{56}$Ni)=$10^{-6} $M$_{\odot}$ can be used. 
The light curves for zero-metallicity models are compared with several RSG models with zero amount of $^{56}$Ni.

\subsubsection{Plateau luminosity}

During the plateau the light curve is powered by shock conversation of the kinetic energy into the thermal energy during the shock propagation in the stellar envelope. The peak luminosity increases with higher explosion energy and radius, but decreases with higher ejecta mass. The absolute V magnitude can be estimated from \citet{Litvinova1985}:
\begin{equation}
V = -2.34 \lg E - 1.80 \lg R + 1.22 \lg M - 11.307 \,, 
\label{eqnV}
\end{equation} 
where $E,R,$ and $M$ are the explosion energy (in 10$^{50}$ erg), the presupernova radius and the ejecta mass in solar units, respectively.

The calculations of zero metallicity models presented
in Figure \ref{z0bolometric} are consistent with the analytic estimations. The plateau of the 25 M$_{\odot}$ and 40 M$_{\odot}$ supernovae having compact progenitor is rather faint: $M_{\rm{bol}}\sim$ -15, while  much larger radius of 100 M$_{\odot}$ progenitor leads to the increase of luminosity. In contrast to the zero metallicity 25-40 M$_{\odot}$ models, solar metallicity SNe are about an order of magnitude more luminous (Figure \ref{rsg}) due to larger radius of the progenitors.

All the above are valid for the models with low amount of $^{56}$Ni. If the $^{56}$Ni mass in the ejecta is rather high ($M$($^{56}$Ni)$>$ 0.01...1M$_{\odot}$), the $^{56}$Ni decay can form more luminous peak. From Figure \ref{z0bolometric} we see that $^{56}$Ni peak is a possible indicator C-normal EMP stars.

The optical multicolor light curves (Figures \ref{ubvri}-\ref{ubvri2}) during the plateau phase demonstrate that the flat shape of the U and B band light curves is similar to the bolometric light curve. The shape of the light curves in more luminous R and I bands is not so flat and has a peak in the middle of the plateau.

\subsubsection{Plateau duration}

The duration $\Delta t$ of the light-curve plateau is estimated by \citet{Litvinova1985} as:
\begin{equation}
\Delta t = -0.191 \lg E + 0.186 \lg R + 0.566 \lg M + 1.047 \,.
\label{eqnT}
\end{equation}

In accordance with this equation, the larger radius and mass provide longer duration of the plateau phase, but larger explosion energy reduces the plateau duration. This behavior is well reproduced in our numerical calculations: in bolometric light curves of zero metallicity (Figure \ref{z0bolometric}) and solar metallicity (Figure \ref{rsg}) models and in multicolor light curves of zero metallicity supernova and hypernova models (Figures \ref{ubvri}-\ref{ubvri2}).

The width of the plateau depends also on the depth of the mixed hydrogen-rich layer \citep[see, e.g.,][]{ShigeyamaNomoto1990}. To estimate the influence of mixing we constructed models with uniform abundances distribution and made a comparison with non-mixed models (Figure \ref{mix}). The mixing does not greatly change the form of the light curve  because the minimum velocity of the hydrogen layer is low for the models.

The solar metallicity RSG progenitor has larger radius which leads to higher luminosity and longer duration of the SN plateau in comparison with BSG progenitors. The 20 M$_\odot$ and 25 M$_\odot$ RSG models have quite a large amount of $^{56}$Ni to make the plateau duration longer in comparison with the low-$^{56}$Ni models (Figure \ref{rsg}). 

\begin{figure}
\centering
  \begin{minipage}[htp]{0.48\textwidth}
  \centering
\includegraphics[width=80mm]{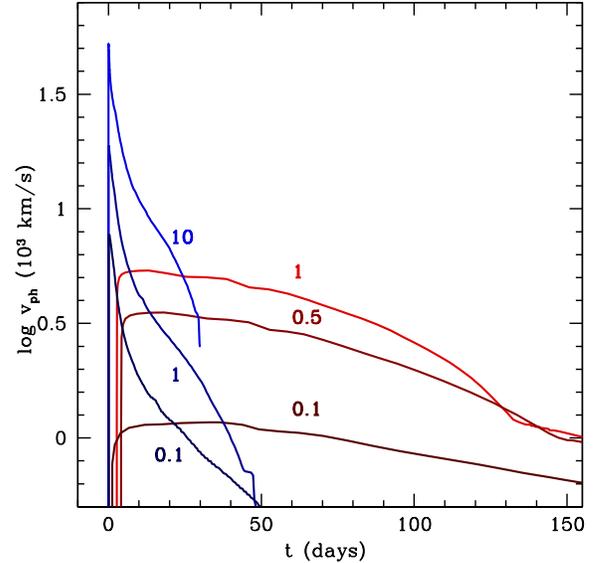}
\caption{Photosperic velocities for 25 M$_{\odot}$ zero metallicity (blue color) and solar metallicity (red color) progenitors with $M$($^{56}$Ni)=0, parametrized by the explosion energy $E_{51}$.} 
\label{faintVel}
\end{minipage}
\end{figure}

\begin{figure}
\centering
  \begin{minipage}[htp]{0.48\textwidth}
  \centering
\includegraphics[width=80mm]{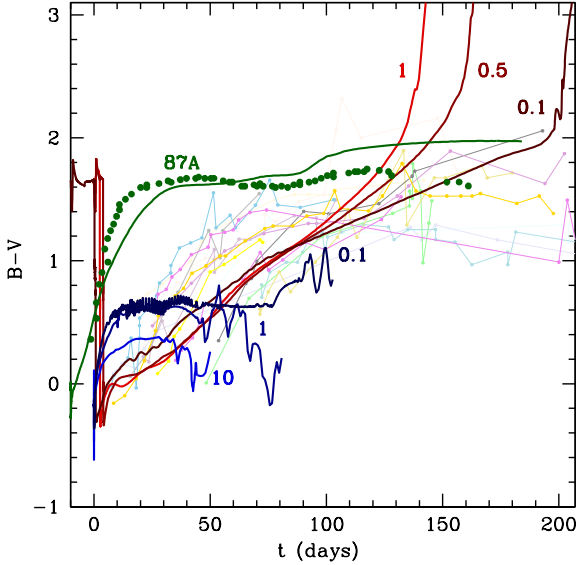}
\caption{Calculated $B-V$ color evolution curves (mag)  
of 25 M$_\odot$ zero-metallicity progenitors (solid blue lines) with $M$($^{56}$Ni)=0, solar metallicity progenitors (solid red lines) with $M$($^{56}$Ni)=0, parametrized by the explosion energy ($E_{51}$=0.1;0.5;1;10) and SN1987A (solid green line). Green dots denote SN 1987A observed $B-V$ color evolution curve \citep{Hamuy1990}, light color lines - the observational data for a set of 16 SN II-P \citep{Hamuy2001}. For calculation of SN1987A color curve (green line) the results of \citet{Blinnikov2000} are used.} 
\label{faintColor}
\end{minipage}
\end{figure}

\begin{figure}
  \begin{minipage}[htp]{0.48\textwidth}
\centering
\includegraphics[width=80mm]{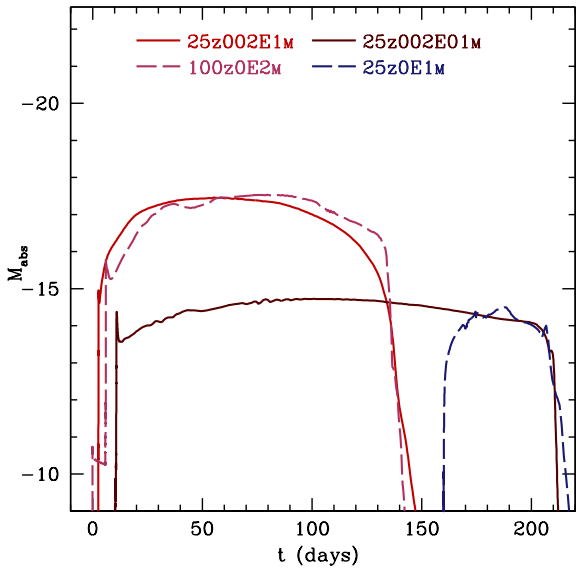}
\caption{V-band light curves for 25 M$_{\odot}$ solar metallicity progenitors with the explosion energy $E_{51}$=0.1;1 (solid brown and red lines), and their comparison with V-band light curves for 25 M$_{\odot}$ ($\bm{E_{51}}$=1) and 100 M$_{\odot}$ ($\bm{E_{51}}$=2) zero metallicity progenitors (dashed blue and maroon colors). Light curve of 25 M$_{\odot}$ zero metallicity model is shifted by 160 days from the moment of explosion. For all the models $\bm{M}$($\bm{^{56}}$Ni)=0.}
\label{faintCmpRSGMix3}
\end{minipage}
\end{figure}

\subsubsection{Photospheric velocities}

In Figure \ref{faintVel} we compare the photosperic velosities for 25 M$_\odot$ zero and solar metallicity models during plateau phase. Due to compact presupernova, zero metallicity model have higher velocity and much faster drop of the photospheric velocity. For hypernova models (explosion energy $E_{51}$=10) the velocity of the outer layers of the ejecta reaches $\sim$ 50,000 km s$^{-1}$. The modeling of the color
The maximum photospheric velocity is realized at the beginning of the plateau phase just after shock breakout epoch.
 
\subsubsection{Color evolution curves}
\label{subsec:Colors}

Figure \ref{faintColor} presents $B-V$ color evolution curves of 25 M$_{\odot}$ progenitors for various values of explosion energy and metallicity. The color evolution curves reveal the common feature: for zero and low metallicity BSG progenitors the $B-V$ value during plateau phase is almost constant, while for solar metallicity RSG progenitors we can see a gradual reddening.

For more general investigation, in Figure \ref{faintColor} we compare the color evolution curves calculated by {\sc stella} with a set of SN~II-P and SN1987A observational data. The modeling of the color evolution curves during the plateau phase demonstrates the cooling rate consistent with observed SNe: the decrease of metallicity leads to flattening of the color evolution curve. The flattening of the color curve also can be seen in observations of low metallicity supernovae \citep[see, e.g.,][]{Polshaw2015}.

It is difficult to use the luminosity and the duration of the plateau phase to distinguish between solar and low metallicity SNe (Figure \ref{faintCmpRSGMix3}). Zero metallicity normal energy SNe are as luminous as solar metallicity low energy SNe. Moreover, the light curve of the massive 100 M$_{\odot}$ zero metallicity SN is quite similar (in the plateau duration and luminosity) to the light curve of less massive (25 M$_{\odot}$) solar metallicity star for the same explosion energy $E_{51}$ = 1. The decline of the luminosity after the plateau phase is also quite similar between the zero and solar metallicity progenitors (Figure \ref{faintCmpRSGMix3}). In this situation the SN color evolution curve can help to distinguish the metallicity of the progenitor.

\begin{figure}
\centering
  \begin{minipage}[htp]{0.48\textwidth}
\centering
\includegraphics[width=80mm]{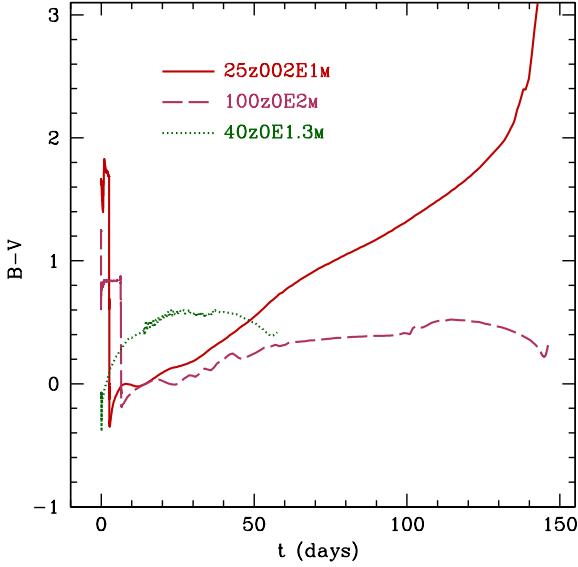}
\caption{Calculated $B-V$ color evolution curves (mag) during plateau phase of 25 M$_\odot$ solar metallicity RSG progenitor (solid red line) with the explosion energy $\bm{E_{51}}$=1, 40 M$_\odot$ zero metallicity BSG progenitor (dotted green line) ($\bm{E_{51}}$=1) and 100 M$_\odot$ zero metallicity RSG progenitor (dashed maroon line) ($\bm{E_{51}}$=2). For all the models $\bm{M}$($\bm{^{56}}$Ni)=0.}
\label{faintColor2}
\end{minipage}
\end{figure}

In Figure \ref{faintColor2} we compare the color evolution curves for zero and solar metallicity RSG models and find that they also differ in the cooling rate and, consequently, can be distinguished in metallicity. For massive zero metallicity BSG presupernova model ({\sc 40z0E1.3}), in contrast to RSGs, the plateau is shorter, but again low metallicity  leads to the flattening of the color evolution curve.

In addition, we confirm the above metallicity dependence with 100 M$_{\odot}$ models in which the metallicity of hydrogen
envelope changes gradually from zero to the solar value. The result of this numerical experiment is the same as above: for zero metallicity model $B-V$ and $U-B$ color evolution curve are flat, but the increase of 
metallicity to the solar value leads to the linear behavior. The color evolution is mostly due to opacity difference between zero and solar metallicity models, while hydrodynamical evolution for both models is similar. 
In distinguishing between zero and solar metallicity models, the evolution of red and infrared color indexes is less informative in comparison with $B-V$ color index.

The different behaviour of low metallicity and solar metallicity color evolution curves can be used to distinguish low metallicity and normal metallicity SNe. 

\begin{figure}
\centering
  \begin{minipage}[htp]{0.48\textwidth}
  \centering
  \includegraphics[width=80mm]{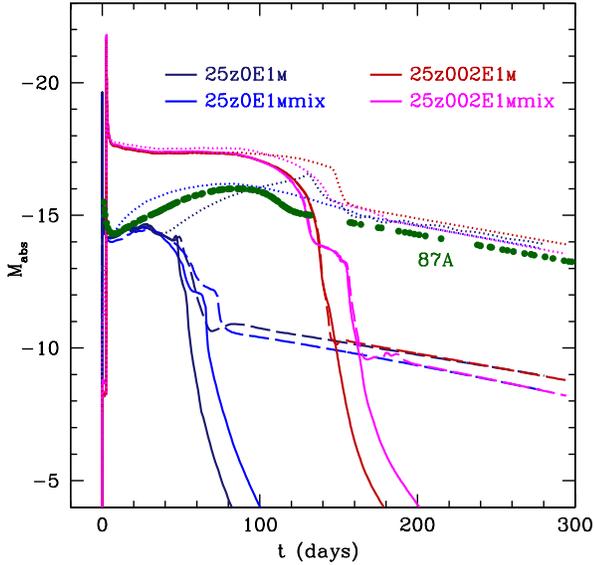}
\caption{Bolometric light curves for zero and solar metallicity mixing-fallback 25 M$_\odot$ supernova models (the explosion energy $E_{51}$ = 1). Blue and magenta colors denote the models with uniform mixing throughout the ejecta.  Solid line -- $M$($^{56}$Ni)=0 (HMP), dashed line -- $M$($^{56}$Ni)=$10^{-3}$ M$_\odot$ (CEMP), dotted line -- $M$($^{56}$Ni)=$10^{-1}$ M$_\odot$ (EMP). The green dots denote bolometric light curve of SN 1987A with $M$($^{56}$Ni)=0.07 M$_\odot$ \citep{SuntzeffBouchet1990}}
\label{faintRSGMix}
\end{minipage}
\end{figure} 

  \begin{figure}
\centering
  \begin{minipage}[htp]{0.48\textwidth}
  \centering
\includegraphics[width=80mm]{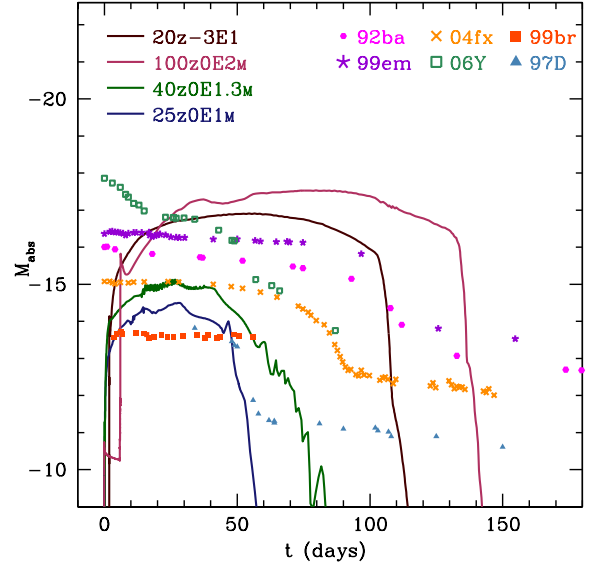}
\caption{V-band light curves for zero metallicity (red, green, blue colors), non-zero metallicity (dark red color) and a number of observed SNe. For all the models $\bm{M}$($\bm{^{56}}$Ni)=0.}
\label{faintCmpRSG}
\end{minipage}
\end{figure}

\subsection{Transition from plateau to $^{56}$Co decay}

After the plateau phase for low $^{56}$Ni models the luminosity starts to decline. The decline rate is defined by the explosion energy \citep{Arnett1980}, opacity, $^{56}$Ni and H mixing \citep[see, e.g.,][]{KasenWoosley2009,Bersten2011,Nakar2015}. 

UBVRI light curves for low $^{56}$Ni models are summarized in Figure \ref{ubvri}-\ref{ubvri2}. The radius of BSGs is smaller in comparison with RSGs and the temperature drops faster \citep{Grasberg1971, Arnett1980}. This leads to the steeper luminosity decline in optical bands for BSG models.

The presence of $^{56}$Ni finally leads to the tail of the light curve powered by the $^{56}$Co decay and its temporal behaviour is given by \citep[see, e.g.,][]{Nadyozhin1994}:
\begin{equation}
M_{\rm{bol}} = -19.19 -2.5\lg{\Big(\frac {M_{\rm{Ni0}}}{M_{\odot}}\Big)} -1.09\frac{t}{\tau_{\rm{Co}}} \,,
\label{eqnM}
\end{equation}
where $t$ is measured from the moment of explosion ($t=0$), $M_{\rm{Ni0}}$ is the total mass of $^{56}$Ni at $t=0$ which decays with a half-life of 6.1 days into $^{56}$Co, and $\tau_{\rm{Co}}$=111.3 days.

The smooth radioactive tail is reproduced in all our numerical calculations with good accuracy.

If we increase $^{56}$Ni mass from zero in the model, the radioactive tail becomes more luminous and at some $M$($^{56}$Ni) mass value its luminosity at the end of the plateau phase can be comparable with the plateau luminosity. This value can be roughly estimated using equations (\ref{eqnV}-\ref{eqnM}), and for our models varies from $\sim$ 0.1-0.01 M$_{\odot}$ for SNe to $\sim$ 0.1-0.5 M$_{\odot}$ for HNe. For these values the nickel peak can be clearly seen in the light curve.

In comparison with zero metallicity SNe models, for solar metallicity models the brightness of the plateau phase is larger and the mass of $^{56}$Ni for forming the nickel peak is larger: $\gtrsim$ 0.1 M$_{\odot}$.

In Figure \ref{ubvri2} we find that the luminosity of the most luminous model {\sc 100z0E60m} without $^{56}$Ni is higher than for the model with $\bm{M}$($^{56}$Ni)=$\bm{10^{-3}}$M$_{\odot}$ in the transition from plateau to the $^{56}$Ni tail up to $\sim$ 150 days. This behavior is explained by the trapping of thermal photons in the model with $^{56}$Ni where the transparency is lower (due to a larger amount of $^{56}$Fe produced by $^{56}$Ni and $^{56}$Co decays). It takes place until the epoch when radioactive decay overwhelms the emission of thermal photons from the entropy reservoir of the shock heated ejecta of zero $^{56}$Ni model at t $>$ 150 days. Although further, more detailed, study may have some sense to clarify fully the nature of this effect, it is not important for the goals of our investigation. The behavior of light curves in those epochs does not change our conclusions.

\subsubsection{Mixing}

The above estimations for the transition phase do not take mixing into consideration, but account for mixing can significantly change the light curve.

To investigate the mixing effect more accurately we compare a number of 25 M$_{\odot}$ models with zero and solar metallicity, varying $M$($^{56}$Ni) and calculating light curves for all these models with uniform mixing throughout the ejecta (Figure \ref{faintRSGMix}). If $M$($^{56}$Ni) is rather large ($\gtrsim$ 0.1 M$_{\odot}$) the mixing reduces rise time to the $^{56}$Ni peak (model {\sc 25z0E1m}). For more plateau luminous models the mixing slightly reduce the plateau duration for the same reason: the $^{56}$Ni peak is shifted to early times (model {\sc 25z002E1m}).

\subsection{Comparison with previous calculations}

We compare the simulation of the multiband light curves of the 25  M$_{\odot}$ hypernova model ($E_{\rm{51}}$ = 10) with light curves calculated by \citet{Smidt2014} for similar models. The luminosity of the plateau and its duration are in good agreement with their calculations. The decline of the light curves after the plateau phase is more sloping in {\sc stella} calculation, which is related to different procedures in opacity calculations. The luminosity at the tail phase must be specified in the future 
in more detailed opacity calculations that include millions of lines, similar to existing procedures for type Ia SNe (private communication with E. Sorokina). Detailed opacity calculations and the analysis of the transition phase (from the plateau to $^{56}$Co decay) in the models with $M$($^{56}$Ni) can help to reveal the physical properties of the presupernova: explosion energy, mixing and opacity.

\section{Observational properties of Pop III and metal-poor core-collapse supernovae}

\subsection{Comparison with the nearby supernovae}
\label{subsec:Comparison}

In Figure \ref{faintCmpRSG} the calculated low-$^{56}$Ni light curves are qualitatively compared with several observed massive Type II SNe (SN 1999em \citep{Elmhamdi2003}, SN 1999br \citep{Pastorello2004},
SN 1997D \citep{Benetti2001}, SN 2004fx \citep{Hamuy2006}, SN 1992ba, SN 2006Y \citep{Anderson2014}).

The luminosity of the plateau for the observed faint supernova SN 1997D is close to the modelled light curves of SNe having BSG progenitors. Adjusted for the uncertainty of the observed plateau duration and possible asymmetry features, our BSG models with low metallicity could be  good candidates for simulation of these object in addition to already existing simulations \citep{Turatto1998,Chugai2000}. 

The B-V color evolution curves of SN 1999em and SN 1999br are included in the set of SN~II-P data \citep{Hamuy2001} presented in Figure \ref{faintColor}. Their color curves are typical to solar metallicity progenitors and the presupernovae of SN 1999em and SN 1999br are supposed to have RSG progenitors. The B-V color evolution curve for SN 1992ba and SN 2004fx 
are also typical to solar metallicity SNe \citep[][]{Jones2009,Hamuy2006}. 

Since we do not know the moment of explosion in observed SN~II-P, it is difficult to distinguish between the low energy explosion of RSG ($E_{51}\sim$ 0.1) and the ordinary BSG SN in the analysis of bolometric light curves (see Figure \ref{faintCmpRSGMix3}). The analysis of photosperic velocities and color curve evolution during the plateau phase can provide more information about the progenitor.

\begin{figure}
\centering
  \begin{minipage}[htp]{0.48\textwidth}
\centering
\includegraphics[width=80mm]{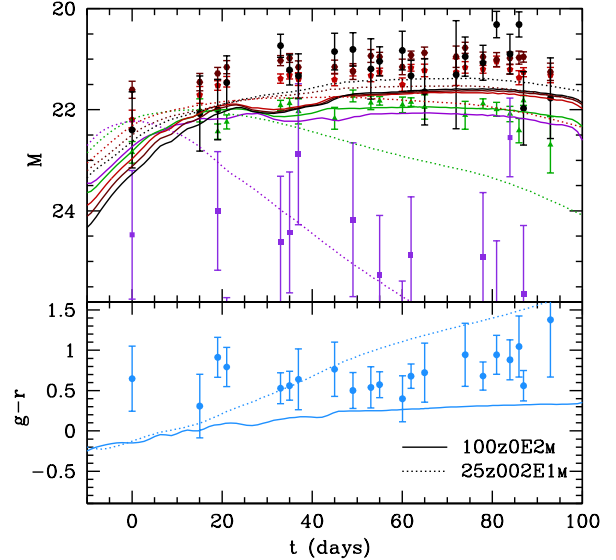}
\caption{Optical SDSS ugriz light curves (upper panel) colour-coded violet, green, red, dark red and black respectively, and color evolution curve (lower panel) for SNII (SDSS ID 12991). Observations are compared with zero metallicity {\sc 100z0E2m} model (solid line) and solar metallicity {\sc 25z002E1m} model (dashed line). For all the models $\bm{M}$($\bm{^{56}}$Ni)=0.} 
\label{id12991}
\end{minipage}
\end{figure}

\begin{figure}
\centering
  \begin{minipage}[htp]{0.48\textwidth}
  \centering
\includegraphics[width=80mm]{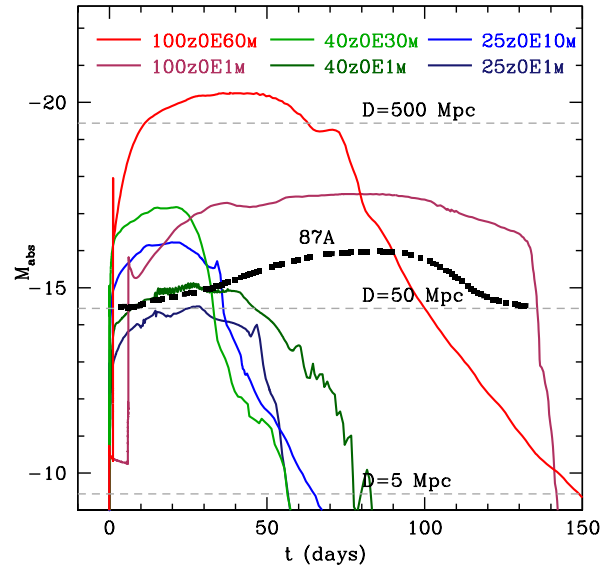}
\caption{V-band light curves for zero metallicity (red, green and blue colors) supernovae with $M$($^{56}$Ni)=0. Horizontal lines represents the Gaia servey detection limits. The dots denote the SN 1987A data of \citet{Hamuy1990}.} 
\label{gaia}
\end{minipage}
\end{figure}

 \begin{figure}
\centering
  \begin{minipage}[htp]{0.48\textwidth}
\centering
\includegraphics[width=80mm]{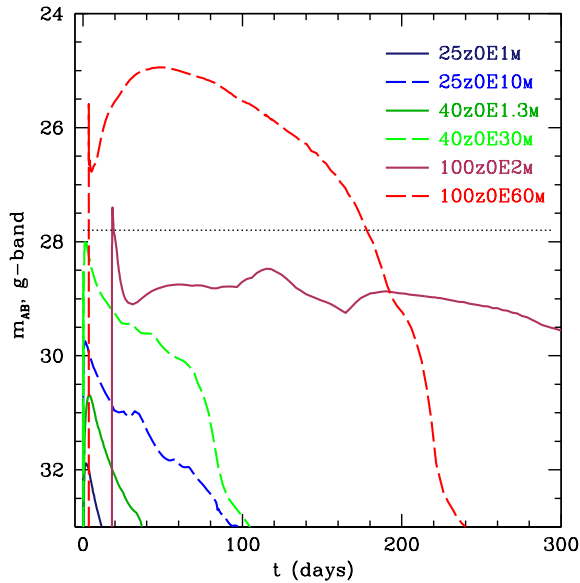}
\caption{Apparent g-band light curves for zero metallicity models at  redshift z=2. No extinction and no IGM absorption are assumed. The horizontal line shows a 5$\sigma$ detection limit in the g-band for the Subaru/HSC 1 hr integration. For all the models $\bm{M}$($\bm{^{56}}$Ni)=0.} 
\label{subaruZ2}
\end{minipage}
\end{figure}

\subsection{Detectability of Pop III supernovae}

While Pop III SNe have typically fainter and shorter plateau than SNe
with solar metallicity, the most important characteristics is the color
evolution curves. The color evolutions at the plateau phase and at the
first $10$~days distinguish the Pop III SNe from RSG SNe with solar metallicity
and SN~1987A-like BSG SNe, respectively (Figure~\ref{faintColor2}). In order to obtain the
color evolution, the multicolor observations should be performed every
$\sim2$ days for the first $10$~days and every $\sim10$~days for the
plateau.

Although the follow-up observations of nearby SNe realize such
multicolor observations with cadences of $<10$~days, SNe with the flat
color evolution have not been discovered (Section \ref{subsec:Comparison}). This might be
because most of previous SN surveys targeted large nearby
galaxies with metal enrichment. Untargeted multicolor SN surveys were performed by SDSS
\citep{Sako2014} and SNLS \citep{Astier2006,Guy2010}. Their cadences are
high enough to draw the color evolution curves at the plateau. We
have checked the multicolor light curves of the SDSS
sample.\footnote{http://data.sdss3.org/sas/dr10/boss/papers/supernova/}
Although the signal-to-noise ratio is poor, an SN~II spectroscopically identified (ID 12991) shows blue color and
flat color evolution curves at the plateau phase (Figure \ref{id12991}). Although
the metallicity of host galaxy is not metal-poor according to $R_{23}$
method ($12 + \log ({\rm O/H}) = 8.58$, \citep{Tremonti2004,Alam2015}),
we propose that the SN took place at the low-metallicity
environment. We do not consider red and infrared color evolution curves because it is not informative to distinguish zero and solar metallicity (see Section \ref{subsec:Colors}).

There are two directions to detect Pop III SNe in low metallicity environment that produce C-normal EMP, CEMP, and HMP stars. 
In order to identify low-$^{56}$Ni SNe, it is required to follow the
light curve until the $^{56}$Co decay. As the tail of faint SNe for the
HMP stars is as faint as $M_{\rm abs}>-7$, the tail can be detected
only if they take place at $<50$~Mpc even with $8$m-class
telescopes. In contrast, their plateau are as bright as
$m_{\rm bol}\sim20$ and thus the detection can be made with an
untargeted survey with such $1$m-class telescopes as PTF \citep{Rau2009}, KISS \citep{Morokuma2014} and the all-sky
 survey as Gaia with the broad-band G band limiting magnitude
 of $G\sim19$ (Altavilla et al. 2012). In Figure \ref{gaia} we plot light curves for low-$^{56}$Ni models
 and designate the detection limits for Gaia mission. For these surveys, EMP star
forming galaxies also can be a good target for finding and identification of supernovae of metal-poor progenitor stars \citep{Pilyugin2014,Thuan2008,Guseva2015}

\begin{figure}[!t]
\centering
  \begin{minipage}[htp]{0.48\textwidth}
\centering
\includegraphics[width=80mm]{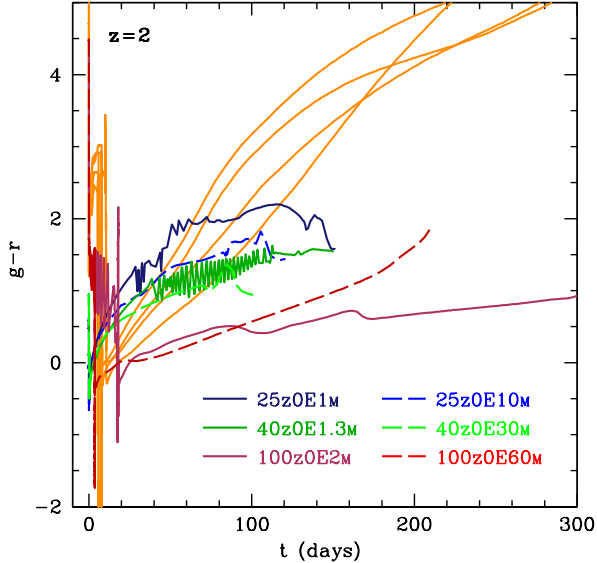}
\caption{Color evolution light curves at redshift z=2 during plateau phase for solar metallicity (orange color) and zero metallicity (blue color - 25 M$_{\odot}$, green color - 40 M$_{\odot}$  and red color - 100 M$_{\odot}$) models. SN models (solid line) and HN models (dashed line) are presented. No extinction and no IGM absorption are assumed. For all the models $\bm{M}$($\bm{^{56}}$Ni)=0.}
\label{faintgrizycolor_2}
\end{minipage}
\end{figure}

The other targets to detect Pop III SNe are the metal-free pockets
at $z\sim2$. Figure~\ref{subaruZ2}
shows the $g$-band light curves of Pop III SNe at $z=2$. Since Pop
III SNe are fainter than SNe with solar metallicity, only the SNe with
$100~$M$_\odot$ can be observed with such $8$m-class telescopes as
Subaru/HSC and LSST. The
explosions with $E_{51}=1$ and $60$ can be detected at the shock
breakout phase and the phase from the shock breakout to the plateau,
respectively. The UV color of Pop III SNe also display the flat color
evolution, in contrast to the increasing UV color evolution of
solar metallicity SNe (Figure \ref{faintgrizycolor_2}).
Therefore, the multicolor optical observation can
inform the metallicity of the SN, although the spectroscopic follow-up
observation is difficult.

\subsection{Contribution to the ionization}

The sources of UV radiation capable to reionize the Universe are still under discussion. According to cosmological models \citep{BarkanaLoeb2001} and observations (\citet{Fan2006}) the reionization of the Universe appears
to have well completed at redshift $z \sim 6$. The first generation stars could be the imporant contributors to the ionization. According to our calculations SNe of RSG progenitors provide larger amount of UV photons at the shock breakout epoch, but still less than BSGs do during lifetime. 

To estimate the contribution of Pop III SNe to cosmic reionization we calculate the number of ionizing photons with the energy $E_{\rm{ion}}=h\nu_{\rm{HI}}>13.6$ eV for all the models:
\begin{equation}
N_{\rm{UV}} = \int_{\nu_{\rm{HI}}}^{\infty} \frac{F_{\nu}}{h\nu} d\nu
\end{equation}

The results are summarized in the last column of Table \ref{modelTableOutput}. The UV photons are mostly produced by shock breakout when the effective temperature is high enough ($T \sim 5 \times 10^5$ K), being similar to previous estimations from the SN 1987A model \citep{LundqvistFransson1996}. Supernovae with larger explosion energies or larger progenitor radii emit larger number of UV photons. But for BSG progenitors the number of UV photons during explosions corresponds only 10 -- 100 yr of the main-sequence lifetime
 \citep{TumlinsonShull2000}. RSG progenitors provides a larger amount of UV photons at shock breakout, but still less than those from BSGs during lifetime. 
 
\section{CONCLUSIONS}
\label{sec:conclusion}

We have calculated the light curves for a number of hydrodynamical models for Pop III 25 -- 100 M$_{\odot}$ core-collapse SNe ($E_{51}\sim$ 1) 
and HNe ($E_{51}\sim$ 10). These models are assumed to undergo mixing and fallback to produce nucleosynthesis yields being well fitted to the observed abundance patterns of EMP, HMP, and CEMP stars.

The radiation-hydrodynamical simulations reproduce in details shock breakout, plateau phase and radioactive tail of the light curves. The observations of shock breakout and multicolor light curves of the plateau phase is important for the identification of zero and low metallicity SNe.

BSGs are typical presupernova for Pop III core collapse SNe with $M\lesssim$ 40-60 M$_{\odot}$ and their structure determine the properties of shock breakout: shorter duration and lower luminosity in comparison with more massive RSG progenitors. The plateau phase is common to both BSG and RSG models and can provide the information that the progenitors are similar to SN~II-P. But the duration of the plateau phase is often unknown from the observation. The evolution of photospere's velocity and multiband light curves can be more useful for the identification of Pop III SNe. We have found that the flat color evolution curve $B-V$ during plateau phase can be used as an indicator of low-metallicity SNe.

The low amount of $^{56}$Ni to explain CEMP stars with mixing-fallback  leads to a sharp luminosity decline after the plateau phase. This feature also can be used as an indicator of low metallicity progenitor. The transition phase from the plateau to the tail can give us the additional information about the explosion energy, mixing and opacity of the presupernova, but it requires more accurate theoretical consideration of opacities in numerical simulations. 

We have modelled Pop III SNe with 1D simulations. The aspherical effects are taken into account approximately with the mixing-fallback model. The mixing of H and $^{56}$Ni could have a large impact on the shape of the light curve and future multi-dimensional radiation calculations are preferred to investigate the effects of asphericity more accurately.

The direct detection of Pop III core-collapse SNe is hardly possible at high redshift \citep{Whalen2013}, but Pop III hypernovae will be visible to the JWST (James Webb Space Telescope) at z $\,\sim\,$ 10 -- 15 \citep{Smidt2014}. The probability of the detection iof Pop III SNe in metal-free gas pockets (z $\,\sim\,$ 2) would be higher, where the detection would be possible by current surveys (HSC/Subaru).

Along with Pop III SNe the results our modeling would be suitable for identification of low-metallicity supernovae in the nearby Universe. The BSG progenitors are supposed to have the metallicity up to 
Z $\,\sim\,$ 10$^{-5}$ -- 10$^{-4}$. There is a number of galaxies in the local Universe with metallicities close to these values \citep{Papaderos2008,Pilyugin2014} and with account of inhomogeneous galaxy regions there would be a good chance for identification and study of these objects. The number of discovered faint supernovae is increasing \citep{Nomoto2012} and the new surveys LSST, JWST (James Webb Space Telescope) are planned to make a large contribution to the detection of low metallicity supernovae. 

\acknowledgments

We thank Hanindyo Kuncarayakti (University of Chile) for the help with providing us the observed SNe data. Also we are grateful to Elena Sorokina (SAI MSU) for useful discussions.

This research is supported by the World Premier International Research
Center Initiative (WPI Initiative), MEXT, Japan and the Grant-in-Aid for Scientific Research of the JSPS (23224004, 26400222), Japan. The work of S.~Blinnikov on modifications of the  code for shock breakout was supported by RNF (grant \textnumero 14-12-00203).

\bibliographystyle{apj}
\bibliography{bibfile}

\end{document}